# Leveraging Quantum Annealing for Large-Scale Household Energy Scheduling with Hydrogen Storage

Arash Khalatbarisoltani[1], Amin Mahmoudi[2], Jie Han[3], Muhammad Saeed[4], Wenxue Liu[1], Jinwen Li[1], Solmaz Kahourzade[5], Amirmehdi Yazdani[6], Xiaosong Hu[1]

**Abstract**— Hydrogen integration into microgrids facilitates the absorption of intermittencies from renewable energy resources. However, significant challenges remain due to complex optimization problems, particularly in large-scale applications involving multiple fuel cells (FCs) and electrolyzers (ELs) with numerous binary decision variables. This paper presents a hierarchical quantum annealing (QA) model predictive control-based power allocation framework aimed at accelerating these optimization problems. First, in a day-ahead stage, the framework determines the startup and shutdown of the FCs and ELs. The short-term stage then refines the output power of the FCs and the hydrogen generation rate of the ELs. The feasibility is evaluated through a case study consisting of multiple households in Australia. Our findings demonstrate that while the traditional optimization approach performs satisfactorily in scenarios with a small number of households, the QA approach becomes more appropriate and effectively solves the problem within an acceptable range as the number of connected households increases.

## 1. Introduction

Worldwide concerns about global warming are prompting the integration of distributed renewable energy (DRE) sources, such as solar photovoltaic panels (PVs) and wind turbines (WTs), into microgrids. Intermittency is one of the main challenges associated with DREs; consequently, conventional microgrids are evolving to incorporate multi-storage systems that enhance reliability while accommodating DREs. Each microgrid with DREs can balance power demand and supply in either interconnected or islanded modes [1]. In [2], the integration of solar PV and battery energy storage for residential houses is investigated, considering vehicle-to-home operation. Another study discusses the effects of electricity tariffs and energy sharing in grid-connected houses [3]. Islanded microgrids that utilize electrolyzers (ELs) and fuel cells (FCs) can benefit from hydrogen ($H_2$) as an energy vector, offering the distinct advantage of long-term storage capabilities that support a stable power supply. An additional benefit is that $H_2$ conversion by-products can also address the water and heating needs of connected facilities, such as residences and hospitals in islanded modes. However, the proliferation of $H_2$ technologies in islanded systems faces challenges due to complex optimization problems and high initial costs. In hybrid power systems (H2PS) that integrate $H_2$ storage, power allocation strategies (PASs) play a crucial role in enhancing performance. A PAS coordinates the operation of various components, such as PEM fuel cells (PEMFCs), ELs, batteries, and DREs, to achieve optimal power allocation under varying uncertain demands. PASs can be categorized into two types: rule-based and optimization-based. Rule-based PASs rely on predefined conditions to guide the operation of H2PS under specific scenarios. For example, a rule-based PAS is proposed in [4] for a standalone H2PS that incorporates WTs and PVs. Furthermore, [5] suggests that FC-based combined heat and power systems can significantly reduce $CO_2$ emissions, while the integration of FCs with supercapacitors is also discussed in [6]. Although rule-based PASs are relatively straightforward to implement, they often do not achieve optimal performance. In contrast, optimization-based PASs leverage predictive information, enabling adaptation to uncertain conditions. However, traditional optimization techniques face scalability limitations when applied to large-scale systems. The integration of FCs and ELs, due to their power constraints, necessitates high computational capacity, further exacerbating the scalability issue. Quantum computing (QC) presents a promising solution [7] and can be divided into two categories: gate-based and annealing-based. In the gate-based category, qubits are initialized in specific states, then manipulated using quantum gates to construct circuits. For example, [8] employs a Quantum Approximate Optimization Algorithm (QAOA) framework based on the Alternating Direction Method of Multipliers (ADMM) to address the unit commitment problem. Despite the benefits associated with these methods, challenges remain when applying them to power systems [9]. Conversely, quantum annealing (QA), in the second category, has demonstrated potential for effectively solving combinatorial optimization problems. For instance, [10] shows that QA outperforms QAOA in various optimization tasks. This study proposes a QA-based model predictive control (MPC) approach for optimal household power management with hydrogen storage, aiming to expedite the optimization solution. This approach utilizes long-term predictions to determine the status of the EL and FC after adjusting their outputs during short-term operations. The remainder of the paper is organized as follows: Section 2 presents the studied standalone households in Australia. Section 3 outlines the QA-based MPC scheduling framework. Section 4 presents the results, followed by Section 5, which summarizes the findings.

[1]Chongqing University, Chongqing, China
[2]Flinders University, Adelaide, Australia
[3]Loughborough University, Loughborough, England
[4]Shanghai Jiao Tong University, Shanghai, China
[5]University of South Australia, Adelaide, Australia
[6]Murdoch University, Perth, Australia

## 2. System model

Fig. 1 presents a multi-standalone household equipped with ELs and hydrogen storage tanks, local residential PEMFCs, and battery packs. The simulation is based on real data obtained from WTs and solar PVs. Data on load demand from one standalone household is available for only one year [11], while similar load profiles are generated for the other households. The operation of the controllable appliances is randomly distributed throughout the year [1]. The output power of each solar PV module, denoted as $P_t^{PV}$, is calculated as [12]

$$P_t^{PV} = \eta_p P_r^{PV} I_c (1 - 0.004(T_a - 25)) \quad (1)$$

where $T_a$ represents the ambient temperature, $I_c$ and $P_r^{PV}$ represent the solar insolation and rated output power of the solar PV module, and $\eta_p$ stands for the solar PV converter efficiency. The output power $P_t^{WT}$ of a WT is considered a piecewise function of wind speed $v$ [12]

$$P_t^{WT} = \begin{cases} 0 & v < v_c \text{ or } v > v_f \\ P_r^{WT} \left(\dfrac{v - v_c}{v_t - v_c}\right)^3 & v_c \leq v < v_t \\ P_r^{WT} & v_t \leq v \leq v_f \end{cases} \quad (2)$$

where $v_c = 3$ m/s, $v_t = 8$ m/s and $v_f = 22$ m/s represent the cut-in, rated and cut-out wind speeds, respectively, and $P_r^{WT}$ denotes the rated power of the WTs.

## 3. Multi-stage QA-based MPC PAS framework

This study proposes a multi-stage scheduling framework to optimize the interconnected households, which integrates DRE sources with ELs and $H_2$ storage, as well as residential PEMFCs and local batteries, as shown in Fig. 2. The multi-stage optimization framework consists of two steps: long-term and short-term. The long-term scheduling step relies on predictions of the DERs and residential load demands. This stage primarily aims to optimize the startup and shutdown statuses of the PEMFCs and ELs for the subsequent 24-hour period using a 1-hour optimization interval, denoted by $T_D$. The objective function of the long-term planning step includes the costs associated with the FCs, ELs, hydrogen storage levels, and battery storage levels. The indices $t$ and $i$ represent the value at time $t$ and the FC and battery of household $i$.

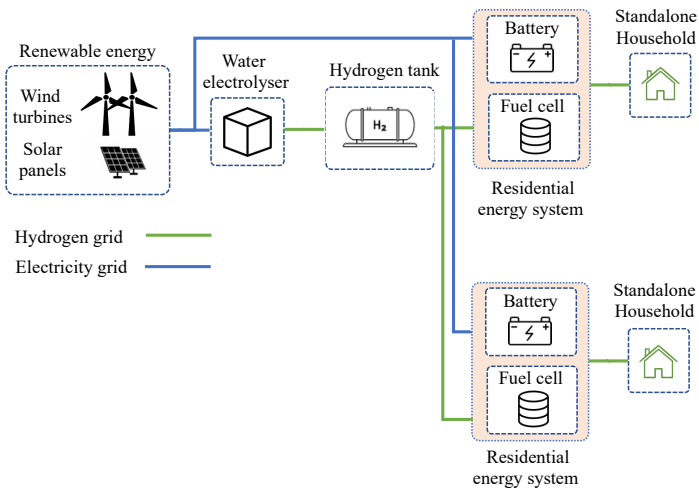

**Fig. 1.** Multi-standalone household integrated with hydrogen storage and renewable energy sources.

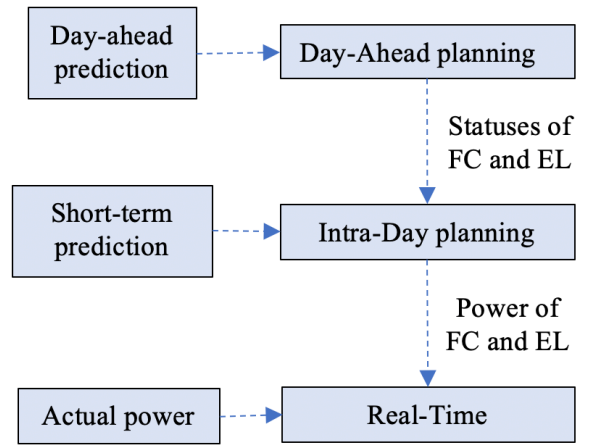

**Fig. 2.** The proposed multi-stage QA-based MPC PAS framework.

$$\min \sum_{i \in I} \sum_{t \in T_D} (J_{i,t}^{FC} + J_{i,t}^{EL}) - H_t + (SOC_t^{bat} - 0.7)^2 \quad (3)$$

$$U_{i,t}^{FC} + S_{i,t}^{FC} + D_{i,t}^{FC} = 1, \quad (4)$$
$$U_{i,t}^{EL} + S_{i,t}^{EL} + D_{i,t}^{EL} = 1, \quad (5)$$
$$U_{i,t}^{FC} = U_{i,t-1}^{FC} + \alpha_{i,t}^{FC} - \beta_{i,t}^{FC}, \quad (6)$$
$$U_{i,t}^{EL} = U_{i,t-1}^{EL} + \alpha_{i,t}^{EL} - \beta_{i,t}^{EL}, \quad (7)$$
$$S_{i,t}^{FC} = S_{i,t-1}^{FC} + \alpha_{s,i,t}^{FC} - \beta_{s,i,t}^{FC} - (\alpha_{i,t}^{FC} - \beta_{i,t}^{FC}), \quad (8)$$
$$S_{i,t}^{EL} = S_{i,t-1}^{EL} + \alpha_{s,i,t}^{EL} - \beta_{s,i,t}^{EL} - (\alpha_{i,t}^{EL} - \beta_{i,t}^{EL}), \quad (9)$$
$$D_{i,t}^{FC} = D_{i,t-1}^{FC} + \alpha_{s,i,t}^{FC} - \beta_{s,i,t}^{FC}, \quad (10)$$
$$D_{i,t}^{EL} = D_{i,t-1}^{EL} + \alpha_{s,i,t}^{EL} - \beta_{s,i,t}^{EL}, \quad (11)$$
$$U_{i,t+k}^{FC} \geq \alpha_{i,t}^{FC}, \quad (12)$$
$$U_{i,t+k}^{EL} \geq \alpha_{i,t}^{EL}, \quad (13)$$
$$D_{i,t+k}^{FC} \geq \beta_{i,t}^{FC}, \quad (14)$$
$$D_{i,t+k}^{EL} \geq \beta_{i,t}^{FC}, \quad (15)$$
$$P_{i,t}^{FC} \geq P_{min}^{FC} U_{i,t}^{FC}, \quad (16)$$
$$P_{i,t}^{EL} \geq P_{min}^{EL} U_{i,t}^{EL}, \quad (17)$$
$$P_{i,t}^{FC} \leq P_{max}^{FC} U_{i,t}^{FC}, \quad (18)$$
$$P_{i,t}^{EL} \leq P_{max}^{EL} U_{i,t}^{EL}, \quad (19)$$
$$SOC_{i,t}^{bat} = SOC_{i,t-1}^{bat} + \frac{\eta_c^{bat} P_{i,t-1}^{ch}}{C^{bat}} - \frac{P_{i,t-1}^{dis}}{\eta_d^{bat} C^{bat}}, \quad (20)$$
$$SOC_{i,t}^{bat} \geq SOC_{min}^{bat}, \quad (21)$$
$$SOC_{i,t}^{bat} \leq SOC_{max}^{bat}, \quad (22)$$
$$y_t^{ch} + y_t^{dis} \leq 1, \quad (23)$$
$$P_{i,t}^{ch} \leq P_{max,c}^{bat} y_t^{dis}, \quad (24)$$
$$P_{i,t}^{dis} \leq P_{max,d}^{bat} y_t^{dis}, \quad (25)$$
$$H_t = H_{t-1} + \eta^p P_{i,t}^{EL} - \eta^c P_{i,t}^{FC}, \quad (26)$$
$$H_t \geq H_{min}, \quad (27)$$
$$H_t \leq H_{max}, \quad (28)$$
$$P_t^{PV} + P_t^{WT} + P_{i,t}^{FC} + P_{i,t}^{EL} + P_{i,t}^{bat} = P_{i,t}^{load}, \quad (29)$$

where $T_D$ denotes the set of hourly intervals across a day. The main objective function comprises costs from PEMFCs and ELs. FCs costs include standby losses and the costs of starting up and shutting down, assuming no losses when the FCs are On and Off [21]:

$$J_{i,t}^{FC} = c_s^{FC} \cdot S_{i,t}^{FC} + c_h^{FC} \cdot (\alpha_{s,i,t}^{FC} + \beta_{s,i,t}^{FC}) + c_l^{FC} \cdot (\alpha_{i,t}^{FC} + \beta_{i,t}^{FC}), \quad (30)$$

Similarly, EL costs cover the costs for standby, and both high and low transition states of ELs and are calculated:

$$J_{i,t}^{EL} = c_s^{EL} \cdot S_{i,t}^{EL} + c_h^{EL} \cdot (\alpha_{s,i,t}^{EL} + \beta_{s,i,t}^{EL}) + c_l^{EL} \cdot (\alpha_{i,t}^{EL} + \beta_{i,t}^{EL}), \quad (31)$$

where $c_s^{FC}$ and $c_s^{EL}$ represent the costs associated with keeping PEMFCs and ELs in standby mode, respectively. $c_h^{FC}$ and $c_h^{EL}$ denote the transition costs related to the startup and shutdown of PEMFCs and ELs, while $c_l^{FC}$ and $c_l^{EL}$ represent the transition costs associated with their operation. Equations (4) and (5) ensure that each PEMFC and EL is in exactly one state (On, Standby, or Off). The binary variables $U_{i,t}^{FC}$, $S_{i,t}^{FC}$, and $D_{i,t}^{FC}$ indicate the On, Standby, and Off statuses of the $FC_i$, respectively. Similarly, $U_{i,t}^{EL}$, $S_{i,t}^{EL}$, and $D_{i,t}^{EL}$ indicate the On, Standby, and Off statuses of the $EL_i$. Equations (6) and (7) define the state transition logic for the On status, linking it to previous states and startup/shutdown actions. Equations (4) and (5) also govern the transitions for the Standby and Off statuses for both $FC_i$ and $EL_i$. These equations impose the enforced hot-start constraints, mitigating the effects of cold starts for PEMFCs and ELs. It is assumed that startup and shutdown actions cannot occur simultaneously. The minimum duration constraints in (6) and (7) enforce that units must operate for a specified minimum duration after starting up or shutting down. $T_{on,min}^{FC}$ and $T_{off,min}^{FC}$ denote the minimum number of time periods a fuel cell must remain On after startup and Off after shutdown, respectively. Similarly, $T_{on,min}^{EL}$ and $T_{off,min}^{EL}$ represent the corresponding minimum durations for the electrolyzer. Equations (16) and (19) impose the minimum and maximum power limits for PEMFCs and ELs. It is assumed that power references are static, and transient characteristics of power are not considered in this phase. $P_{i,t}^{FC}$ and $P_{i,t}^{EL}$ represent the power output of the $FC_i$ and the power consumption of the $EL_i$, respectively. $P_{i,min}^{FC}$ and $P_{i,max}^{FC}$ indicate the minimum and maximum allowable output powers of the $FC_i$, while $P_{i,min}^{EL}$ and $P_{i,max}^{EL}$ denote the minimum and maximum allowable power consumptions of the $EL_i$. The battery state of charge (SOC), denoted as $SOC_{i,t}^{bat}$, is governed by Equation (20), which tracks changes in SOC over time. Equations (21) and (22) enforce the minimum and maximum SOC limits. The SOC dynamically changes based on stored energy, and charging and discharging efficiencies are represented by $\eta_c^{bat}$ and $\eta_d^{bat}$, respectively, although these values may differ from one another. $P_{(i,t-1)}^{ch}$ and $P_{(i,t-1)}^{dis}$ are the power charged into and discharged from the battery at time $t-1$, while $C^{bat}$ is the battery capacity. $SOC_{min}^{bat}$ and $SOC_{max}^{bat}$ are the minimum and maximum allowable states of charge for the battery, respectively. To prevent simultaneous charging and discharging, Equation (23) is applied, while Equations (24) and (25) enforce the charging and discharging power limits. $y_t^{ch}$ and $y_t^{dis}$ are binary variables indicating whether the battery is charging or discharging. $P_{max,c}^{bat}$ and $P_{max,d}^{bat}$ are the maximum power limits for charging and discharging the battery. Hydrogen storage changes over time are modeled in Equation (26), with Equations (27) and (28) ensuring that the storage remains within capacity limits. The efficiencies for hydrogen production and consumption by ELs and PEMFCs are also considered as different constants. $H_t$ represents the amount of hydrogen stored, while $\eta^{EL}$ and $\eta^{FC}$ are the efficiencies for hydrogen production and consumption, respectively. $H_{min}$ and $H_{max}$ are the minimum and maximum hydrogen storage capacities, respectively. Equation (29) enforces the power balance constraint, where $P_t^{PV}$ is the power generated by solar PVs, $P_t^{WT}$ is the power generated by WTs, $P_{i,t}^{FC}$ is the power output from the PEMFC, $P_{i,t}^{EL}$ is the power consumed by the EL (considered as negative generation or load), and $P_{i,t}^{load}$ is the total electrical load demand. The short-term MPC-based stage optimizes the power outputs of the PEMFCs and ELs. The main objective is to minimize costs related to output fluctuations, denoted as $C_f$, while maximizing the stored $H_2$ and maintaining the $SOC$ level of residential batteries, under the previous mentioned constraints.

$$\min \sum_{t \in T_{pred}} -H_{2,t}^{Str} + C_{f,t} + (SOC_t^{bat} - 0.7)^2 \tag{32}$$

$$C_{f,t} = \sum_{i \in I} \sum_{t \in T_P} (|\Delta P_t^{FC}| + |\Delta P_t^{EL}|), \tag{33}$$

where $T_P$ denotes the 15-minute planning horizon, distinct from the 1-hour long-term scheduling period $T_D$ considered in the previous stage. $\Delta P_t^{FC}$ and $\Delta P_t^{EL}$ represent the power changes for the PEMFCs and ELs, respectively. To implement on the QA platform, the optimization problem is formulated as a Quadratic Unconstrained Binary Optimization (QUBO) problem and then transformed into the Ising model [13]. To ensure that the transformation's accuracy loss is limited, constraints are integrated into the QUBO as penalty terms [14, 15]. The Hamiltonian of the optimization problem is constructed using the spin variables obtained from the discrete operational variables. This Hamiltonian formulation is minimized using the QA technique. The Hamiltonian formulations of all constraints and objective functions are not detailed; however, two examples are presented, and the rest follow the same analogous principles. As an example, the PEMFC state is encoded by

$$h_s^{FC} = \sum_{t \in T} \alpha_1 \left(\frac{1+f_t^W}{2} + \frac{1+f_t^V}{2} - 1\right)^2, \tag{34}$$

where $\alpha_1$ denotes the penalty coefficient, and $f_t^W$ and $\xi_t^V$ represent the spin variables related to the states of the PEMFC, respectively. Similarly, the PEMFC maximum and minimum output power constraints are applied by

$$h_P^{FC} = \sum_{t \in T} \alpha_2 \left(P_{min}\frac{1+f^u}{2} - \frac{1+\bar{f}^u}{2} + s_1\right)^2 + \alpha_3 \left(\frac{1+\bar{f}^u}{2} - P_{max}\frac{1+f^u}{2} + s_2\right)^2, \tag{35}$$

where $\alpha_2$ and $\alpha_3$ are penalty terms, $s_1$ and $s_2$ represent non-negative slack variables, and $\bar{f}^u$ is the discretized and spin-transformed power output of the PEMFC.

## 4. Results and discussions

This section evaluates the performance of the hierarchical QA-based MPC PAS framework for multiple households with hydrogen storage, compared to the Gurobi-based approach. The QA-based implementation is achieved through D-Wave's API, which uploads the problem to a remote QA platform [13]. The values for $SOC_{min}$ and $SOC_{max}$ are set at 20% and 90%, respectively, with an initial SOC of 60%. Each household comprises one residential customer equipped with a 2 kW PEMFC and a 2 kW EL. The hydrogen storage tank has a capacity of 15 kg, with an initial value of 1 kg. Fig. 3 presents the results obtained from the proposed method for one household, effectively demonstrating the advantages of employing both long-term and short-term strategies. The PEMFC is utilized primarily during daytime hours when renewable energy generation is expected to meet load demands. When the battery's SOC is low, the PEMFC supplements the power load as needed. Fig. 4 illustrates the operation of the PEMFC and EL over a one-year profile using the proposed QA-based approach. At the beginning of the year, external renewable energy contributes to an increase in hydrogen storage; however, this stored hydrogen is later consumed. These results clearly show the advantage of hydrogen storage and its ability to store energy for extended periods compared to battery storage, which has limited capacity.

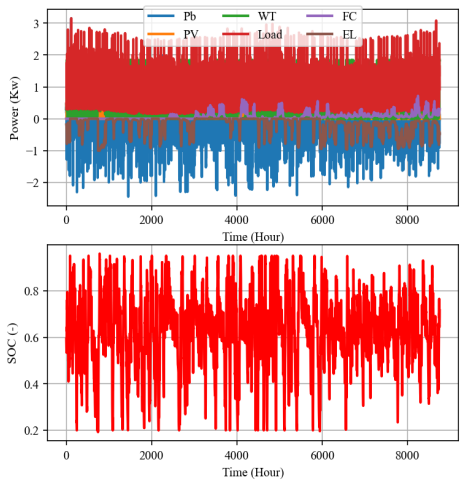
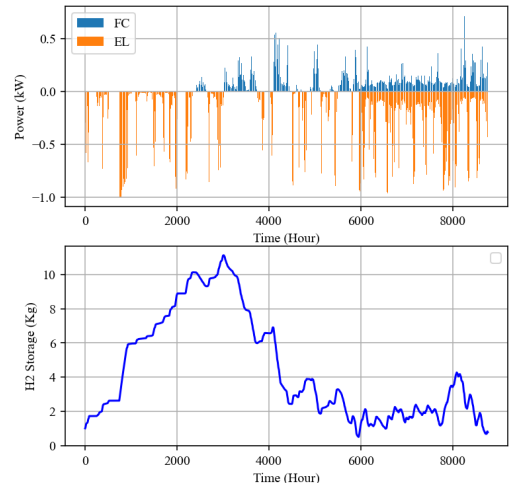

**Fig. 3.** a) power output profiles, b) soc SOC levels.　　　　**Fig. 4.** a) PEMFC and EL power profiles, b) hydrogen storage.

## 5. Conclusion

This paper presents a hierarchical QA-based MPC framework designed for multiple households with storage capabilities. The feasibility of the proposed method was assessed using real data from a case study conducted in Australia. The approach consists of a long-term stage that determines the statuses of the PEMFCs and ELs based on projected energy demands over a one-hour horizon, as well as a short-term stage that adjusts the power outputs of the PEMFCs and ELs over a 15-minute horizon. The QA-based multi-stage strategy enhances the management of storage, preventing the depletion of low-cost resources. Our findings demonstrate that while the traditional optimization approach performs satisfactorily in scenarios with a small number of households, the QA approach becomes more suitable and effectively solves the optimization problem within an acceptable range as the number of connected households increases.